# Feedback to the European Data Protection Board's Guidelines 2/2023 on Technical Scope of Art. 5(3) of ePrivacy Directive

Cristiana Santos, Utrecht University, The Netherlands
Nataliia Bielova, Inria, France
Vincent Roca, Inria, France
Mathieu Cunche, Inria, France
Gilles Mertens, Inria, France
Karel Kubicek, ETH Zurich, Switzerland
Hamed Haddadi, Imperial College London, UK

We very much welcome the EDPB's Guidelines. Please find hereunder our feedback to the Guidelines 2/2023 on Technical Scope of Art. 5(3) of ePrivacy Directive.[1] Our comments are presented after a quotation from the proposed text by the EDPB in a box.

## 1 INTRODUCTION

> 4. The aim of these Guidelines is to conduct a technical analysis on the scope of application of Article 5(3) ePD, namely to clarify what is covered by the phrase '*to store information or to gain access to information stored in the terminal equipment of a subscriber or user*'. These Guidelines do not intend to address the circumstances under which a processing operation may fall within the exemptions from the consent requirement provided for by the ePD.

These guidelines could mention their applicability to member states wherein the ePrivacy regulator is a member of the EDPB, but also to other ePrivacy regulators (non-DPAs), as happens with several other types of regulators.

## 2 ANALYSIS
## 2.1 Key elements for the applicability of Article 5(3) ePD

> 6. Article 5(3) ePD applies if:
>
> d. **CRITERION D:** the operations carried out indeed constitute a '*gaining of access*' or '*storage*'. Those two notions can be studied independently, as reminded in WP29 Opinion 9/2014: *'Use of the words "stored or accessed" indicates that the storage and access do not need to occur within the same communication and do not need to be performed by the same party.*
>
> For the sake of readability, the entity gaining access to information stored in the user's terminal equipment will be hereafter referred to as an 'accessing entity'.

---

[1] Available online at https://edpb.europa.eu/system/files/2023-11/edpb_guidelines_202302_technical_scope_art_53_eprivacydirective_en.pdf.

Since the guidelines cover not only the technical scope of Art. 5(3) but also the general application of this article -- including paragraphs related to legal persons, interplay with the right to privacy, relation to personal data, we suggest editing the title of the guidelines to refer to the general scope of this article.

## 2.2 Notion of 'information'

> 7. As expressed in CRITERION A, this section details what is covered by the notion of 'information'. The choice of the term, much broader than the notion of personal data, is related to the scope of the ePrivacy Directive.

Using the more general term 'information' is welcomed, as elements used for fingerprinting or profiling can be technical[2] (ex.: network identifiers, flags, options, etc) and their qualification as personal data is not straightforward.

> 9. In fact, scenarios that do intrude into this private sphere even without involving any personal data are explicitly covered by the wording of the Article 5(3) ePD and by Recital 24, for example the storage of viruses on the user's terminal. This shows that the definition of the term 'information' should not be limited the property of being related to an identified or identifiable natural person.

We believe that the explanation of coverage of 'storage' and 'gaining access' by the Article 5(3) is ambiguous in the current guidelines - please see our general comment right before Section 3 USE CASES.

> 11. Whether the origin of this information and the reasons why it is stored in the terminal equipment should be considered when assessing the applicability of Article 5(3) ePD have been previously clarified, for example in the WP29 Opinion 9/2014: '*It is not correct to interpret this as meaning that the third-party does not require consent to access this information simply because he did not store it. The consent requirement also applies when a read-only value is accessed (e.g. requesting the MAC address of a network interface via the OS API)*'.

The meaning of stored information should be clarified to specify that it is not limited to data stored in a file system or in memory (as pointed out in paragraph 37). For example, MAC addresses are intrinsically existing on the terminal equipment as being tied to the hardware.

## 2.3 Notion of 'Terminal Equipment of a Subscriber or User'

> 15. Whenever a **device** is not an endpoint of a communication and **only conveys information without performing any modifications to that information**, it would not be considered as the terminal equipment in that context. Hence, if a device solely acts as a **communication relay**, it should not be considered a terminal equipment under Article 5(3) ePD.

---

[2] G. Celosia and M. Cunche, "Saving Private Addresses: An Analysis of Privacy Issues in the Bluetooth-Low-Energy Advertising Mechanism," in *MobiQuitous 2019 - 16th EAI International Conference on Mobile and Ubiquitous Systems: Computing, Networking and Services*, Dec. 2019, pp. 1–10, available online at 10.1145/3360774.3360777.

The status of "communication relays" should be clarified for the following reasons:
- In the case of a router provided by an ISP to a user when it performs Network Address Translation (NAT)[3] which modifies the information transmitted on the network (IP, port, packet checksum), it is unclear if it would enter in the definition of a "communication relay";
- It is unclear whether proxy servers[4] that modify the transmitted information would be covered by the ePD or not;
- It is unclear whether a VPN[5] endpoint, that encrypts/decrypts a whole packet, and thus modifies information, would be covered by the ePD or not;
- Fingerprinting these devices[6] is possible and could be used to recognise users. Such fingerprinting actions should be definitely covered by Article 5(3) ePD and we invite EDPB to clarify such applications.

> 16. A terminal equipment may be comprised of any number of individual pieces of hardware, which together form the terminal equipment. This may or may not take the form of a physically enclosed device hosting all the display, processing, storage and peripheral hardware (for example, smartphones, laptops, connected cars or connected TVs, smart glasses).

We would like to highlight that many upcoming devices do not have displays, such as IoT, Humane's Ai Pin and devices for vision impairment humans, and credit cards. Additionally, we believe the phrase "any number of individual pieces of hardware" should be changed because a virtual terminal equipment, hosted remotely in the context of a software-as-a-service, also constitutes a terminal equipment. For example, a user who needs a specific smartphone app, but has no compatible smartphone, can use a hosting service that gives her an emulated smartphone where she can install/use this app. We believe such virtual terminal equipment should also be covered by Article 5(3).

## 2.4 Notion of 'electronic communications network'

> 20. Another element to consider in order to assess the applicability of Article 5(3) ePD is the notion of 'electronic communications network'. In fact, the situation regulated by the ePD is the one related to *the provision of publicly available electronic communications services in public communications networks in the Community*. It is therefore crucial to delimit the electronic communications network context in which Article 5(3) ePD applies.

The statement above is taken from the Article 3 ePD 2009: "*This Directive shall apply to the processing of personal data in connection with the provision of publicly available electronic communications services in public communications networks in the Community*", that mentions only personal data. This is confusing for the reader within the scope of current guidelines and should be clarified.

---

[3] https://en.wikipedia.org/wiki/Network_address_translation
[4] https://en.wikipedia.org/wiki/Proxy_server
[5] https://en.wikipedia.org/wiki/Virtual_private_network
[6] E. Marechal and B. Donnet, "Network Fingerprinting: Routers under Attack", in *2020 IEEE European Symposium on Security and Privacy Workshops (EuroS&PW)*, Genoa, Italy, 2020, pp. 594-599, doi: 10.1109/EuroSPW51379.2020.00086.

It would be useful to also provide examples of public and private electronic communications services to make a clear distinction for the reader. Similarly, examples of public and, in contrast, private communication networks are very welcome.

> 24. This definition of network does not give any limitation with regards to the number of terminal equipment present in the network at any time. Some networking schemes rely on asynchronous information propagation to present peers in the network and can at some point in time have as little as two peers communicating. Article 5(3) ePD would still apply in such cases, as long as the network protocol allows for further inclusion of peers.

It should be noted that a terminal equipment may transmit information when it is not connected to a network. This is especially the case with wireless network technologies such as Wi-Fi and Bluetooth that employ discovery mechanisms in order to find available devices or networks. With these discovery mechanisms, a terminal equipment can broadcast messages for extended duration.[7]-[8] Thus information on the terminal equipment can be passively collected even if a connection is not established.

> 25. The public availability of the communication service over the communication network is necessary for the applicability of Article 5(3) ePD[9]. It should be noted that the fact that the network is made available to a limited subset of the public (for example, subscribers, whether paying or not, subject to eligibility conditions) does not make such a network private.

It is unclear whether the last sentence of this paragraph is meant to cover public services (with limited number of users/subscribers). It is unclear what would be an example of a communication network with a limited subset of the public. Please also provide an example of a private network and, separately, of a private service, to make it explicit what kinds of networks and services are not therefore covered. For example, is it suggested that when a terminal device used by an employee connects to a company network, Article 5 (3) does not apply?

## 2.5 Notion of 'gaining access'

> 27. In a nutshell, the ePD is a privacy preserving legal instrument aiming to protect the confidentiality of communications and the integrity of devices. In Recital 24 ePD, it is clarified that, in the case of natural persons, the user's terminal equipment is part of their private sphere and that accessing information stored on it without their knowledge may seriously intrude upon their privacy.

---

[7] K. Fawaz, K.-H. Kim, and K. G. Shin, "Protecting Privacy of BLE Device Users," in *25th USENIX Security Symposium (USENIX Security 16),* Austin, TX: USENIX Association, 2016, pp. 1205–1221, available online at https://www.usenix.org/conference/usenixsecurity16/technical-sessions/presentation/fawaz, and

[8] J. Freudiger, "How talkative is your mobile device?: an experimental study of Wi-Fi probe requests," in *Proceedings of the 8th ACM Conference on Security & Privacy in Wireless and Mobile Networks*, ACM, 2015, p. 8. Accessed: May 27, 2016, available online at http://dl.acm.org/citation.cfm?id=2766517.

We believe "integrity of services" should be replaced by "integrity of terminal equipment" (e.g., see paragraph 14.). Devices are not used to refer to the terminal equipment in this document (e.g., see paragraph 10).

> 28. Legal persons are also safeguarded by the ePD. In consequence, the notion of 'gaining access' under Article 5(3) ePD, has to be interpreted in a way that safeguards those rights against violation by third parties.

It is unclear whether this statement means that storage of information or access to information on a device owned by a private company is also covered by Article 5(3) ePD.

> 29. Storage and access do not need to be cumulatively present for Article 5(3) ePD to apply. The notion of 'gaining access' is independent from the notion of 'storing information'. Moreover, the two operations do not need to be carried out by the same entity.

We posit that it is unclear in the current guidelines, whether the mere storage of information alone is covered by the Article 5(3) ePD. Please see our general comment right before Section 3 USE CASES.

> 31. Whenever the accessing entity wishes to gain access to information stored in the terminal equipment and actively takes steps towards that end, Article 5(3) ePD would apply. Usually this entails the accessing entity to proactively send specific instructions to the terminal equipment in order to receive back the targeted information. For example, this is the case for cookies, where the accessing entity instructs the terminal equipment to proactively send information on each subsequent HTTP (Hypertext Transfer Protocol) call.

In case of browser cookies, in general, no active steps are needed from the accessing entity to gain access to the cookies because cookies that were previously stored in the browser are automatically sent to the entity thanks to the 'Cookie HTTP request' header.[9] The paragraph above would be correct if an example of browser storage, such as localStorage, was taken instead of cookies, since localStorage can only be actively accessed via a browser API accessing through a JavaScript code provided by the accessing entity.

User's information can also be accessed without any proactive sending. Certain browser fingerprinting features can be collected 'passively'[10] and we encourage EDPB to clarify that such scenarios should still be covered by Article 5(3) ePD. We also believe that device fingerprinting techniques should be more explicitly included in the current guidelines.

The notion of "actively takes steps" should also be clarified pointing the reader to the use case of URL tracking, where the user identifier is contained inside the URL. In this case, it should be clear that by clicking on a URL, the entity gains access to the identifier, even though it was not technically stored on the user's terminal equipment.

---

[9] See https://en.wikipedia.org/wiki/List_of_HTTP_header_fields and https://datatracker.ietf.org/doc/html/rfc6265#section-5.4.

[10] See Table 1 at Imane Fouad, Cristiana Santos, Arnaud Legout, Nataliia Bielova. My Cookie is a phoenix: detection, measurement, and lawfulness of cookie respawning with browser fingerprinting, in *Privacy Enhancing Technologies Symposium* (PoPETS 2022). Available online at https://hal.science/hal-03218403v2/document.

German supervisory authorities have so far provided a different view. The German Conference of Data Protection Authority guidelines[11] only cover "access" to information if this is targeted and thus actively requested from a terminal's storage, under paragraphs 21 and 22.
-Paragraph 21: "*Access requires a targeted transmission of browser information that is not initiated by the end user. If only information, such as browser or header information, is processed that is transmitted inevitably or due to (browser) settings of the end device when calling up a telemedia service, this is not to be considered "access to information already stored in the end device"*. Examples of this are:
• the public IP address of the terminal equipment,
• the address of the called website (URL),
• the user agent string with browser and operating system version and
• the set language.
-Paragraph 22: ''*In contrast, it is already considered access to information on the end user's terminal equipment if the properties of a terminal are actively read - for example, by means of JavaScript code - and transmitted to a server for the creation of a fingerprint.*"

The Guidelines from the Baden-Württemberg DPA[12] states under point 3.1 (page 14) that ''*Section 25 TTDSG only covers "access" to information if it is targeted. Both the IP address and user agent are information that the browser automatically sends when you access a website, without the provider of the telemedia service being able to influence this. The server (unlike a cookie) has not stored any information on the user's device to identify the user and it does not "access" (or initiate access to) information. The information was sent to him without any involvement from the provider of the telemedia service. This procedure is therefore not covered by Section 25 TTDSG.*"

The current draft guidelines need thus to clarify this divergence.

> 33. In some cases, the entity instructing the terminal to send back the targeted data and the entity receiving information might not be the same. This may result from the provision and/or use of a common mechanism between the two entities. For example, one entity may have used protocols that imply the proactive sending of information by the terminal equipment which may be processed by the receiving entity. In these circumstances, Article 5(3) ePD may still apply.

The word "targeted" in the phrase "targeted data" may be confused by a reader with the concept of "targeted content", we therefore propose to rephrase it with "selected data", "data to be shared" or "data from the terminal". We invite EDPB to give concrete examples when Article 5(3) ePD would apply instead of using a vague statement "may still apply".

---

[11] German Conference of Data Protection Authorities (Datenschutzkonferenz) guidelines for "Telemedia Providers" (OH Telemedien 2021), https://www.datenschutzkonferenz-online.de/media/oh/20211220_oh_telemedien.pdf.
[12] Cookies and tracking by website operators and smartphone app manufacturers, 2022, https://www.baden-wuerttemberg.datenschutz.de/wp-content/uploads/2022/03/FAQ-Tracking-online.pdf.

One specific example can be the password manager that auto-fills the email address of the user when the user is visiting a web form. Exfiltration[13] of this email address by an entity present on the web form should fall within the scope of Article 5(3) ePD. We give another example that we invite EDPB to use under paragraph (51).

## 2.6 Notions of 'Stored Information' and 'Storage'

> 34. Storage of information in the sense of Article 5(3) ePD refers to placing information on a physical electronic storage medium that is part of a user or subscriber's terminal equipment[11].

Similarly to our previous comment under paragraph 16, we invite EDPB to consider adding virtual equipment next to the physical electronic storage in this paragraph.

> 36. The ePD does not place any upper or lower limit on the length of time that information must persist on a storage medium to be counted as stored, nor is there an upper or lower limit on the amount of information to be stored.

We welcome the interpretation of storage of any duration (even ephemeral).

> 38. As long as the networked storage medium constitutes a functional equivalent of a local storage medium (including the fact that its only purpose is for the user of the terminal equipment to store information that will be processed on the terminal equipment itself), that storage medium will be considered part of the terminal equipment.

We agree a NAS can be regarded as part of a terminal equipment, but a NAS often hosts many advanced services beyond storage, some of which may also send user information to remote servers. We posit that it is potentially also a terminal equipment per se.

> 39. Finally, 'stored information' may not just result from information storage in the sense of Article 5(3) ePD as described above (either by the same party that would later gain access or by another third party). It may also be stored by the user or subscriber, or by a hardware manufacturer, or any other entity; be the result of sensors integrated into the terminal; or be produced through processes and programs executed on the terminal equipment, which may or may not produce information that is dependent on or derived from stored information.

We would like to comment globally on the notions of 'gaining access' and 'stored information/storage' since the current guidelines introduce inconsistencies. Let us consider the following example: an entity
1) stores 'information' (under the scope of ePD) on the terminal equipment (for example, a user identifier stored in the user's browser storage, such as localStorage or cache),
2) transforms this information (for example, by applying a hashing function to the identifier),
3) and then sends this information to another entity (for example, a JavaScript code sends it to a third party via an HTTP request).

---

[13] Senol, A., Acar, G., Humbert, M., & Borgesius, F. Z. (2022). Leaky Forms: A study of email and password exfiltration before form submission, in *31st USENIX Security Symposium* (USENIX Security 22) (pp. 1813-1830), available online at
https://www.usenix.org/system/files/sec22fall_senol.pdf

See the steps of this process in the figure below.

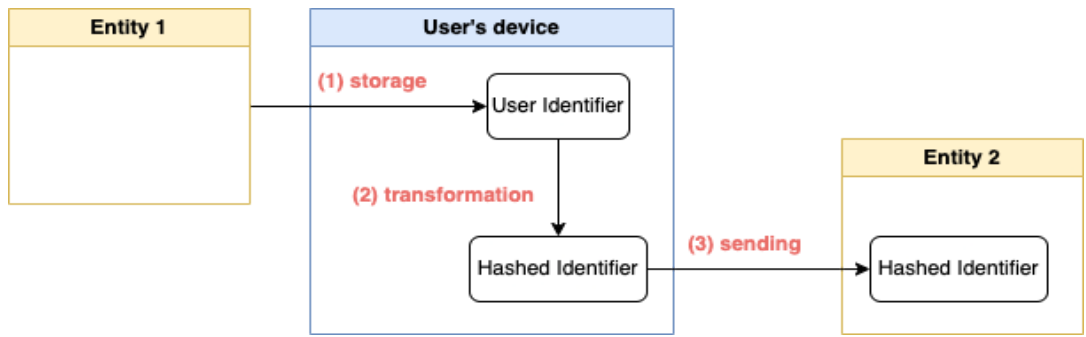

The current guidelines therefore sometimes claim that on-device storage and transformation are covered by Article 5(3) ePD, and in other parts of the guidelines such on-device actions are explicitly not covered:

- Paragraph 9 states that mere storage of information on the user's terminal equipment is covered by the Article 5(3) ePD and by Recital 24, invoking an example that the storage of viruses on the user's terminal is covered. Therefore, in case (3) sending does not occur, (1) and (2) are covered by ePD;
- Paragraph 29 also states that "Storage and access do not need to be cumulatively present for Article 5(3) ePD to apply." Therefore, (1) and (2) should be covered by Article 5(3) ePD even if (3) sending never occurs;
- Paragraph 43 explicitly mentions that use of information strictly inside of the terminal is not subject to Article 5(3) ePD as long as information doesn't leave the device. This means that without the sending operation, (1) and (2) alone are not covered by ePD. Paragraph 53 also refers to a similar scenario and claims that the sending operation corresponds to the 'gaining access' mentioned in Article 5(3) ePD.

Since Article 5(3) explicitly mentions 'the storing of information, or the gaining of access to information already stored', we believe it is important for EDPB to remove this ambiguity from the guidelines and explicitly state which parts of the process (storage, transformation and sending) are indeed covered by Article 5(3).

## 3 USE CASES

> 41. Network communication usually relies on a layered model that necessitates the use of identifiers to allow for a proper establishment and carrying out of the communication. The communication of those identifiers to remote actors is instructed through software following agreed upon communication protocols. As outlined above, the fact that the receiving entity might not be the entity instructing the sending of information does not preclude the application of Article 5(3) ePD. This might concern routing identifiers such as the MAC or IP address of the terminal equipment, but also session identifiers (SSRC, Websocket identifier), or authentication tokens.

The concept of ''receiving entity'' should be clarified. In the context of an explicit connection, the receiving entity is clearly identified (via a destination address). However, in the context of broadcast communication, the receiving entity can be any device in range. It is for instance the case in wireless networks where discovery messages are broadcasted and are intended to be received by all devices in range. It should be specified that the receiving entity can be an entity explicitly identified in the context of the communication, but can also be any entity that might receive the information.

It should be noted that the concerned layers can be as low as the physical layer, where it has been shown[14] that a number of elements can be leveraged to fingerprint devices.

The last phrase of (41) includes "session identifiers" and "authentication tokens", however it was clear from the Article 29 Working Party, "Opinion 04/2012 on Cookie Consent Exemption"[15] that such purposes are exempted of consent[16]. We therefore ask EDPB to clarify how purposes requiring consent and that are exempted from consent interplay with the statements in these guidelines.

> 42. In the same manner, the application protocol can include several mechanisms to provide context data (such as HTTP header including 'accept' field or user agent), caching mechanism (such as ETag or HSTS) or other functionalities (cookies being one of them). Once again, the abuse of those mechanisms (for example in the context of fingerprinting or the tracking of resource identifiers) can lead to the application of Article 5(3) ePD.

We observe that the statement "abuse of those mechanisms" above implicitly relates to the notion of "legitimate use" of such mechanisms. Relating to our previous comment under paragraph 41, we invite EDPB to make it clear, and to specify what purposes of the usage of such mechanisms constitute abuse.

> 43. On the other hand, there are some contexts in which local applications installed in the terminal uses some information strictly inside the terminal, as it might be the case for smartphone system APIs (access to camera, microphone, GPS sensor, accelerator chip, radio chip, local file access, contact list, identifiers access, etc.). This might also be the case for web browsers that process information stored or generated information inside the device (such as cookies, local storage, WebSQL, or even information provided by the users themselves). The use of such information by an application would not be subject to Article 5(3) ePD as long as the information does not leave the device, but when this information or any derivation of this information is accessed through the communication network, Article 5(3) ePD may apply.

The last sentence states that Article 5(3) ePD does not apply as long as information does not leave the device, however this contradicts previous statements in these guidelines. Please see our general comment right before Section 3 USE CASES. Additionally, we invite EDPB to avoid statements like "Article 5(3) ePD may apply" and if this is necessary, then specify under which conditions Article 5(3) ePD will apply in this context.

---

[14] T. D. Vo-Huu, T. D. Vo-Huu, and G. Noubir, "Fingerprinting Wi-Fi Devices Using Software Defined Radios," in *Proceedings of the 9th ACM Conference on Security & Privacy in Wireless and Mobile Networks*, in WiSec '16. New York, NY, USA: ACM, 2016, pp. 3–14, , available online at 10.1145/2939918.2939936.

[15] See https://ec.europa.eu/justice/article-29/documentation/opinion-recommendation/files/2012/wp194_en.pdf.

[16] For a summary overview, see Table 5 at Cristiana Santos, Nataliia Bielova and Célestin Matte. Are cookie banners indeed compliant with the law? Deciphering EU legal requirements on consent and technical means to verify compliance of cookie banners, in *International Journal on Technology and Regulation* (TechReg), 2020, available online at https://techreg.org/article/view/10990/11964.

## 3.1 URL and pixel tracking

> 47. In the case of an email, the sender may include a tracking pixel to detect when the receiver reads the email. Tracking pixels on websites may link to an entity aggregating many such requests and thus being able to track users' behaviour. Such tracking pixels may also contain additional identifiers as part of the link. These identifiers may be added by the owner of the website, possibly related to the user's activity on that website. They may also be dynamically generated through client-side applicative logic. In some cases, links to legitimate images may also be used for the same purpose by adding additional information to the link.

We highlight that a tracking pixel in itself, actually as any other content loaded within the website, cannot track the user if no additional information is sent, such as user identifier encoded in the URL, or sent as a value of the cookie. We also note that any website content may be used to track the user, not only tracking pixels. Indeed, previous research shows that mere loading of JavaScript libraries, big visible images and third-party html content, fonts and stylesheets actually track users with cookie-based techniques.[17]

> 50. Under the condition that said pixel or tracked URL have been distributed over a public communication network, it is clear that it constitutes **storage on the communication network user's terminal equipment, at the very least through the caching mechanism of the client-side software**. As such, Article 5(3) ePD is applicable.

> 51. The inclusion of such tracking pixels or tracked links in the content sent to the user constitutes an instruction to the terminal equipment to send back the targeted information (the specified identifier). In the case of dynamically constructed tracking pixels, it is the distribution of the applicative logic (usually a JavaScript code) that constitutes the instruction. As a consequence, it can be considered that the collection of the identifiers provided by tracking pixels and tracked URL do constitute a 'gaining of access' in the meaning of Article 5(3) ePD and thus the latter is applicable to that step as well.

We invite EDPB to clarify paragraphs 50 and 51 and consider the following example of tracked links. Imagine the following scenario:[18] 1) a social network that creates a unique ID to track Alice who is visiting this social network; 2) this ID is integrated in the URL that Alice sends to Bob over a private messaging application; Bob receives the URL and clicks on it, thus 3) sending the ID of Alice to another entity. This iteration is depicted in the Figure below.

---

[17] See Table 6 at Imane Fouad, Nataliia Bielova, Arnaud Legout, Natasa Sarafijanovic-Djukic. Missed by Filter Lists: Detecting Unknown Third-Party Trackers with Invisible Pixels, in *Privacy Enhancing Technologies* (PETs'20), available online at https://petsymposium.org/popets/2020/popets-2020-0038.pdf

[18] The "tracked URL" technique is called "link decoration", it was recently found that this technique is observed on 73% of visited websites (Section 5.1) of Munir, S., Lee, P., Iqbal, U., Shafiq, Z., & Siby, S. (2023). PURL: Safe and Effective Sanitization of Link Decoration, arXiv preprint available online at https://arxiv.org/abs/2308.03417.

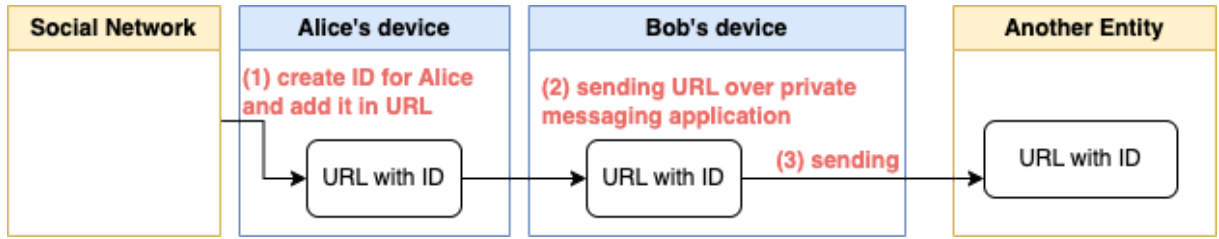

What is unclear in the current guidelines is whether the creation of an ID for Alice and the fact that she has copied the URL constitutes "storage" under Article 5(3). Given that the sending of the URL from Alice to Bob happened over a private messaging application, it is not clear whether this sending (step 2) is covered by Article 5(3).

Finally, we invite the EDPB to consider that the sending (step 3) of Alice's ID to another entity should be covered by Article 5(3) even though there is no active sending of the information; the sending of the information simply happens because Bob has clicked on the URL that Alice has shared with him.

## 3.4 Intermittent and mediated IoT reporting

> 57. Some IoT devices have a direct connection to a public communication network, for example through the use of WIFI or a cellular SIM card. IoT devices might be instructed by the manufacturer to always stream the collected information, yet still locally cache the information first, for example until a connection is available.

We think that: "...through the use of WIFI or cellular SIM card" is inappropriate (e.g., the SIM card does not send anything). We suggest: "...through the use of a Wi-Fi network or a cellular network".

> 58. Other IoT devices do not have a direct connection to a public communication network and might be instructed to relay the information to another device through a point-to-point connection (for example, through Bluetooth). The other device is generally a smartphone which may or may not pre- process the information before sending it to the server.

The wording: "The other device is generally a smartphone" is inappropriate. IoT encompasses a broad range of equipments, from quantified self devices, to smart homes, smart cities, connected vehicles, etc. For sure, many of these devices cannot connect directly to the public network, yet the situation where a smartphone relays information applies to a limited number of use-cases. We suggest instead:

"The other device can be a smartphone (e.g., in case of quantified self trackers), or a dedicated hub (e.g., in case of a connected device in a smart-home, connected through a Wi-Fi, Zigbee or Bluetooth interface to a dedicated hub or smart speaker)."

> 60. In the case of IoT devices connected to the network via a relay device (a smartphone, a dedicated hub, etc.) with a purely point to point connection between the IoT device and the relay device, the transmission of data to the relay could fall outside of the Article 5(3) ePD as the communication does not take place on a public communication network. However, the information received by the relay device would be considered stored by a

> terminal and Article 5(3) ePD would apply as soon as this relay is instructed to send that information to a remote server.

The transmission of data to the relay does not always happen outside a public communication network. Although wireless technologies such as Bluetooth originally transmit data over a 'private' piconet centered around a smartphone, some recent implementations do not rely on a network to transmit data. For instance, in Bluetooth Low Energy, data is transmitted (broadcast) in advertising packets (who will be received by all nearby receivers) outside a network.[19]-[20]

---

[19] A. Heinrich, M. Stute, T. Kornhuber, and M. Hollick, "Who Can Find My Devices? Security and Privacy of Apple's Crowd-Sourced Bluetooth Location Tracking System," *Proceedings on Privacy Enhancing Technologies,* vol. 2021, no. 3, pp. 227–245, Jul. 2021, doi: https://doi.org/10.2478/popets-2021-0045.

[20] G. Celosia and M. Cunche, "Discontinued Privacy: Personal Data Leaks in Apple Bluetooth-Low-Energy Continuity Protocols," *Proceedings on Privacy Enhancing Technologies*, vol. 2020, no. 1, pp. 26–46, Jan. 2020, doi: 10.2478/popets-2020-0003.